# An Efficient Max-Min Resource Allocator and Task Scheduling Algorithm in Cloud Computing Environment

J. Kok Konjaang
Bolgatanga Polytechnic,
Bolgatanga, Ghana

J.Y. Maipan-uku
Ibrahim Badamasi Babangida
University Lapai, Nigeria

Kumangkem Kennedy
Kubuga
Tamale Polytechnic, Ghana

## ABSTRACT
Cloud computing is a new archetype that provides dynamic computing services to cloud users through the support of datacenters that employs the services of datacenter brokers which discover resources and assign them Virtually. The focus of this research is to efficiently optimize resource allocation in the cloud by exploiting the Max-Min scheduling algorithm and enhancing it to increase efficiency in terms of completion time (makespan). This is key to enhancing the performance of cloud scheduling and narrowing the performance gap between cloud service providers and cloud resources consumers/users. The current Max-Min algorithm selects tasks with maximum execution time on a faster available machine or resource that is capable of giving minimum completion time. The concern of this algorithm is to give priority to tasks with maximum execution time first before assigning those with the minimum execution time for the purpose of minimizing makespan. The drawback of this algorithm is that, the execution of tasks with maximum execution time first may increase the makespan, and leads to a delay in executing tasks with minimum execution time if the number of tasks with maximum execution time exceeds that of tasks with minimum execution time, hence the need to improve it to mitigate the delay in executing tasks with minimum execution time. CloudSim is used to compare the effectiveness of the improved Max-Min algorithm with the traditional one. The experimented results show that the improved algorithm is efficient and can produce better makespan than Max-Min and DataAware.

## 1. INTRODUCTION
The concept of utility computing, grid computing and visualization through which cloud computing emerged has brought significant advantages for the development and transformation of businesses, organizations and individuals by providing reliable, customized, quality of service (QoS) and cost beneficial Information Technology (IT) services to business entities, academic institutions, organizations and the general public on demand in daily basis via internet [1]. There are many computer resources in the cloud environment that offer users the opportunity to have access to which includes; processing power, bandwidth, storage, etc. Cloud computing is widely accepted in the world because it offers a varied array of IT services [2] to users with the vision of utility and grid computing, where users access and pay for services offered them in a ways similar to paying for household utilities such as water, telephones, electricity and gas [3], [4].

The idea behind cloud computing is to transform a bigger segment of the IT market, making software and hardware a more attractive service and redefining the use of IT hardware and software for effective resource allocation and scheduling. The rapid growth in the development of cloud systems and applications into the cloud computing environments is a very challenging task [5] and needs efficient algorithms for effective allocation of resources. The current Max-Min algorithm select tasks with maximum execution time on a faster available machine or resources that is capable of giving minimum completion time. The concern of this algorithm is to give priority to tasks with maximum execution time [6], [7] by executing them first before assigning those with the minimum execution time for the purpose of minimizing computational time [8]. The drawback of Max-Min algorithm is that, the execution of tasks with maximum execution time first may increase the total response time of the system [9] and leads to a delay in executing tasks with minimum execution time, hence the need to improve on the current Max-Min algorithm to mitigate the delay in executing tasks with minimum completion time.

### 1.1 Cloud Services Models
Cloud computing is web based technology that provides infrastructure to users as a service (IaaS), platform as a service (PaaS), and software as services (SaaS), [10][7]. SaaS is a cloud service model which is hosted on the server to provide consumers with cloud applications running on its infrastructure. Customers have no control over the management of the services. Examples of SaaS applications are: *content management systems*, *customer relationship management offerings*, *video conferencing* and *e-mails communication systems* [11]. In PaaS, customers are provided with the opportunity to develop and run web applications using programming languages, services, libraries and tools supported by the cloud providers. The consumer has no power of managing or controlling the deplored infrastructure, but has control over the deployed applications and can possibly configure the settings [12]. IaaS is a cloud platform that provides some fundamental computer resources like processing power, memory, networks and other computer resources on demand to customer/users to deploy and run their applications on them. Customers can only run their application on the cloud resources, but have no control over the management of the infrastructures. The three (3) major cloud service models are summarized in figure 1 below.

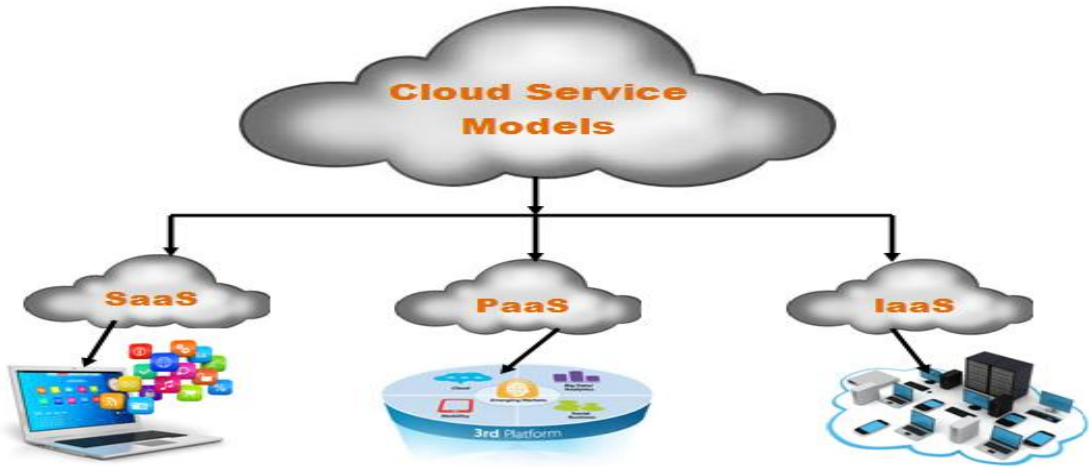

**Figure 1. Cloud computing services model**

This paper considered Infrastructure as a Service (IaaS)

## 1.1 Type of cloud computing

In cloud computing technology, huge amounts of computer resources are connected for the purpose of sharing and information propagation. This can be done through public, private, community and hybrid cloud as elaborated in figure 2.

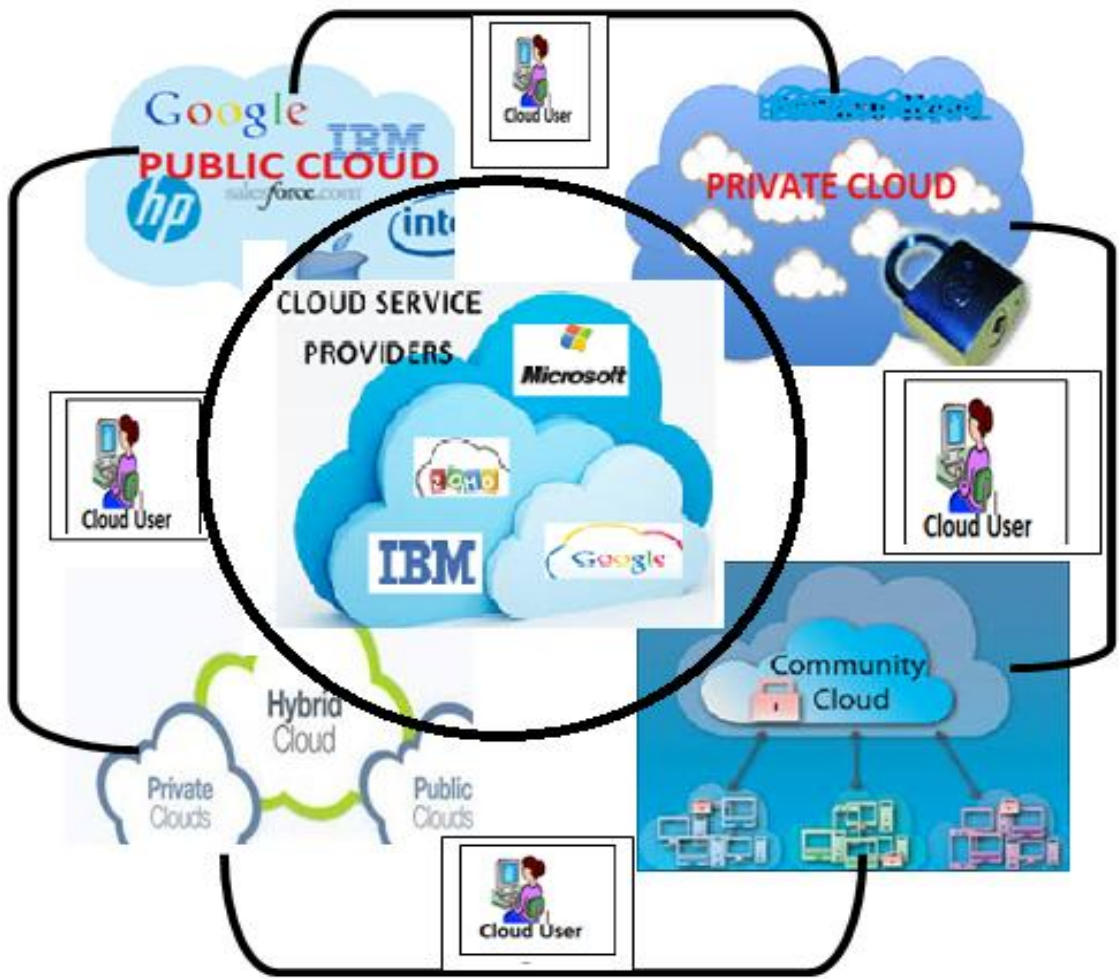

**Figure 2: Types of cloud computing**

### *1.2.1 Public Cloud:*
These are cloud computing services that are made available and accessible to the doorsteps of the broad-spectrum of the public by public cloud service providers who host and maintain the infrastructure. Public cloud may be owned, managed and controlled by academic institutions, business organizations, government organizations or a combination of these. Examples of cloud services aimed at the general public may include online storage services, e-mail services, social networking sites and Apps (such as Facebook, WhatsApp, and Viber) [13]. Figure 3 below depicts the summary of public cloud providers.

### *1.2.4 Hybrid Cloud:*
It is a cloud computing platform that uses a combination of two or more different cloud infrastructure (public, private or community) with unique entities that allows the movement of work loads between the two cloud providers to enable data and application portability. E.g. cloud bursting for load balancing between clouds [13].

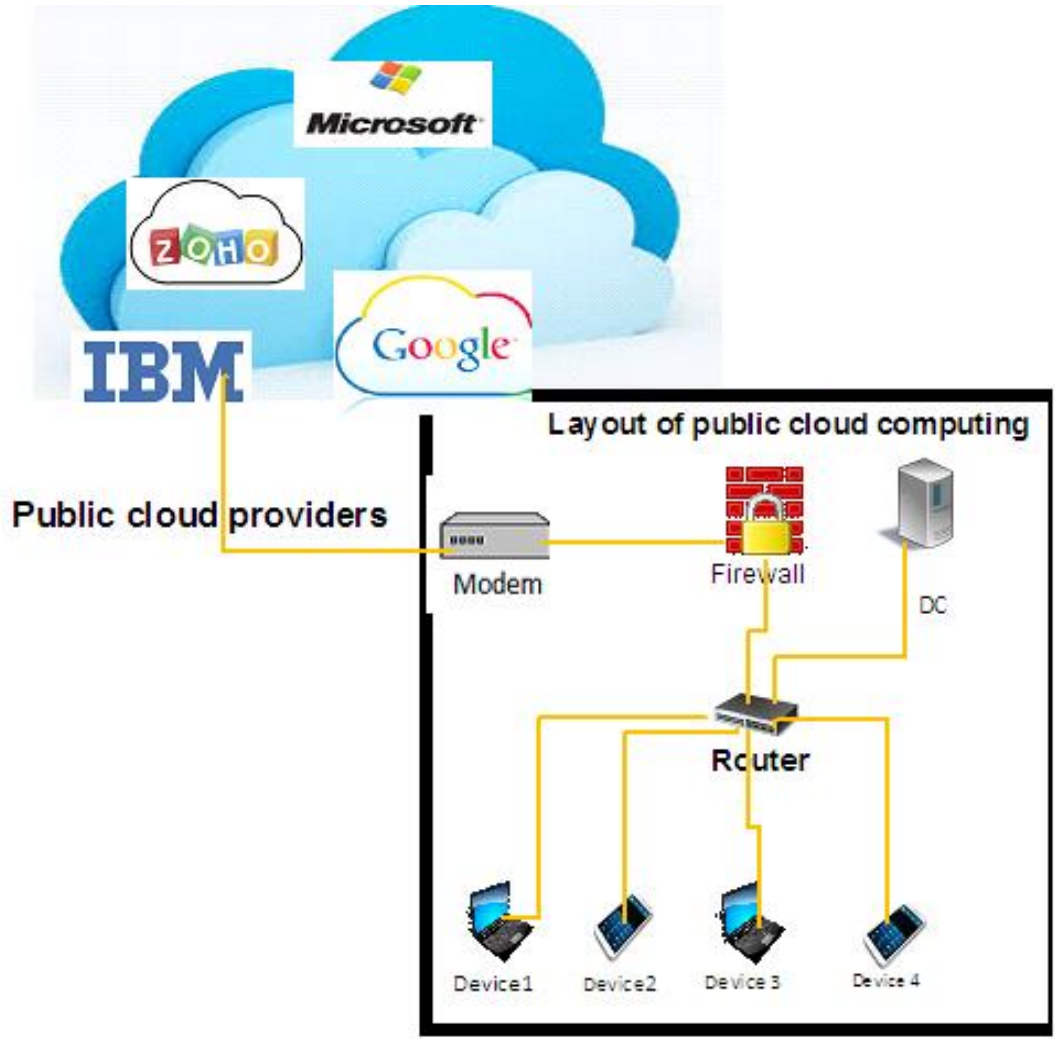

**Figure 3: Layout/Summary of public cloud providers**

### *1.2.2 Private Cloud:*
This is a cloud computing platform that delivers IT as a service rather than a product to customers with a reduced cost. It offers customers higher efficiency with improved innovative business models that are attractive and recommended for the development and transformation of their businesses in this technological era [14]. It is owned, managed and operated by private organizations, third parties or amalgamation of both.

### *1.2.3 Community cloud*
Community cloud is a multi-tenant platform that falls between public and private cloud with respect to their market segment [15] that provide cloud infrastructure to some group of individuals or customers within an organization with a similar interest or requirement (e.g., mission, security requirements, policy, and compliance considerations). In community cloud, the cost of accessing the cloud infrastructure is less expensive as compared to public and private cloud because it is shared among the organizations.

## 2. RELATED WORKS
The major concern in cloud computing is insuring that resources are allocated effectively to cloud applications via the Internet to support cost effective and efficient use of cloud resources [16] for the purpose of minimizing cost and maximizing throughput. Resource allocation is defined by [17] as a way by which distributed resources are assigned and executed on cloud applications via the Internet. To be able to allocate resources effectively, efficiently and just-in-time, efficient cloud resource allocator mechanisms or policies is required to ensure cloud users are satisfied with the services provided [18].

In solving the problem of resource allocation in the cloud, Game theory [19] was studied. In their study, two practical approximated solutions are proposed. A Binary Integer Programming method is introduced to deal with independent optimization, and evolutionary mechanism that will handle multiplexed strategies by minimizing losses. [20] Proposed a Dominant Resource Fairness (DRF) for a fair and sharing model that simplifies Max-Min fairness to multiple resource types. The purpose of the proposed DRF was to satisfy a number of desirable properties and also to incentivize users to share resources by ensuring that users perform at least as well in a shared cluster as they would in smaller and separate clusters for effective resource allocation.

In [9], two algorithms based on improving Max-Min algorithm aimed at assigning tasks by using arithmetic means and geometric means formulas were proposed. In their proposed work, algorithm 1 suggests the use of arithmetic mean to compute average execution time for values that are independent to get the best average execution time, otherwise; the geometric mean should be used for the computation of the best average mean for tasks that depend on other values. Their experimental results show that using arithmetic and geometric means calculations, instead of selecting the job with the highest execution time at all the times, selecting the average size of the task through arithmetic and geometric means calculations can produce a better makespan and average utilization of resources.

## 3. PROPOSED ALGORITHM

THE Improved Max-Min algorithm works in two phases. In phase one like the traditional Max-Min algorithm, It assembles all the cloudlets in increasing order, this means that cloudlets with minimum execution time are in the front and those with the maximum execution time in the rear of the queue. Then a new set (list) of cloudlets using equation 1 is generated and applied Max-Min method for execution.

$$\text{NewcloudletList} = vm / N * \text{cloudletsize} \quad \text{Eq. 1}$$

*where N is the total number of vm.*

$$\text{Completion Time } (CT_{ij}) = et_i + rt_j \quad \text{Eq. 2}$$

Where: $et_i$ is the execution time for cloudlet $c_i$ and $rt_j$ is the ready time of $vm$ $r_j$ [2]. The flowchart is given in figure 3 below.

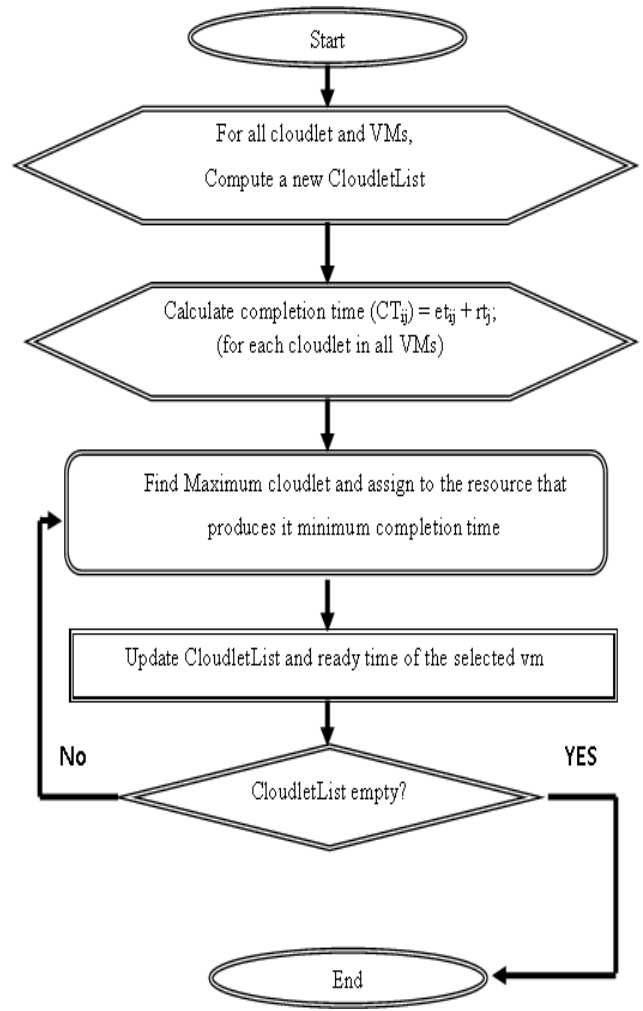

**Figure 4. Flowchart for improved algorithm**

Pseudo Code of Improved Max-Min algorithm
1. While there are cloudlets in CloudletList
2. for all submitted cloudlets in the set; $c_i$
3. for all VMs; $vm_j$
   a. Compute a new CloudletList
   b. Calculate completion time $(CT_{ij}) = et_{ij} + rt_j$; (for each cloudlet in all VMs)
   c. find the cloudlet with maximum execution time (MaxClt)
4. assign MaxClt to the vm that produce it minimum execution time
5. End if
6. update the CloudletList
7. Update ready time $(rt_j)$ of the selected $vmR_j$
8. Update $ct_{ij}$ for all $c_i$
9. End While

## 4. IMPLEMENTATION, EXPERIMENTS AND RESULTS ANALYSIS

To be able to authenticate the proposed algorithm, an experiment was conducted using HP laptop computer with 64 bit windows 8 operating system, Intel (R) core(TM) i3 Processor 2.0GHz, 4G RAM, and 500GB hard drive. We used CloudSim toolkit 3.0.2 simulator. CloudSim toolkit is a simulation environment used for conducting experiments and simulating the behaviour of algorithms in dynamic cloud environments. For simulation purposes, 15 Virtual Machines (VMs) were chosen. We used two different real dataset workflows to represent our data set which includes; Inspiral and CyberShake. We used 50, 100 and 1000 cloudlets for each workflow in all the three algorithms. Two datacenters were created with one datacenter broker who is responsible for discovering, selection and submitting cloudlets to the algorithm for scheduling. The Proposed (Improved Max-Min) algorithm was benchmarked against other two algorithms such as Max-Min and DataAware. The experimental setting is simplified in the table 2 below.

**Table 1.0 workflows and task settings**

| Workflows | Small load | Median Load | Large load |
|---|---|---|---|
| CyberShake | 50 | 100 | 1000 |
| Inspiral | | | |

**Table 2. Parameters setting**

| Description of entities | Qualifications |
|---|---|
| Workflows | Inspiral |
| | CyberShake |
| | 50 |
| | 100 |
| | 1000 |
| Data center | 2 |
| Virtual Machines | 15 |
| Data center broker | 1 |

**Table 3. Makespan comparison using Inspiral**

| | INSPIRAL _50 | INSPIRAL _100 | INSPIRAL _1000 |
|---|---|---|---|
| Improved-Max-Min | 3085.68 | 6510.94 | 27163.01 |
| Max-Min | 3271.68 | 5486.32 | 28124.73 |
| DataAware | 10126.53 | 6703.38 | 28051.93 |

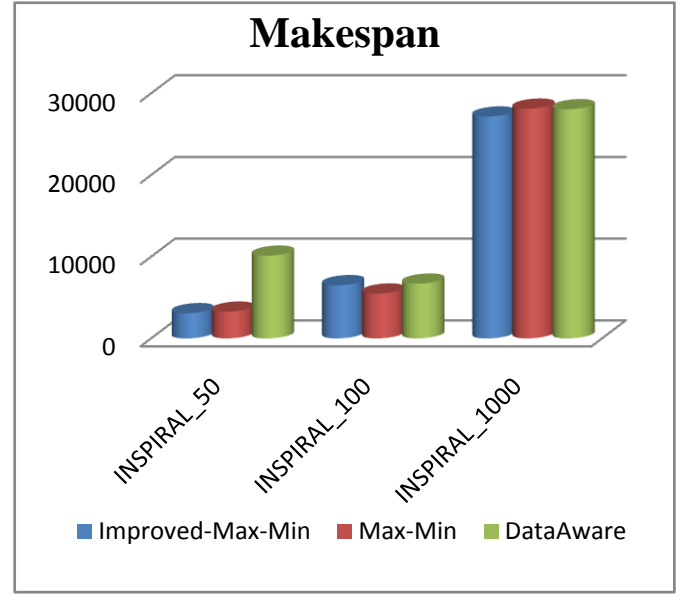

**Figure 5. Makespan Comparison of different algorithms using 50, 100 and 1000 cloudlet of Inspiral**

**Table 4. . Makespan comparison using CyberShake**

| | CYBERSH AKE _50 | CYBERSHA KE _100 | CYBERSHA KE _1000 |
|---|---|---|---|
| Improved-Max-Min | 632.26 | 984.6 | 3287.49 |
| Max-Min | 659.54 | 1023.76 | 3174.45 |
| DataAware | 826.31 | 1157.7 | 2531.15 |

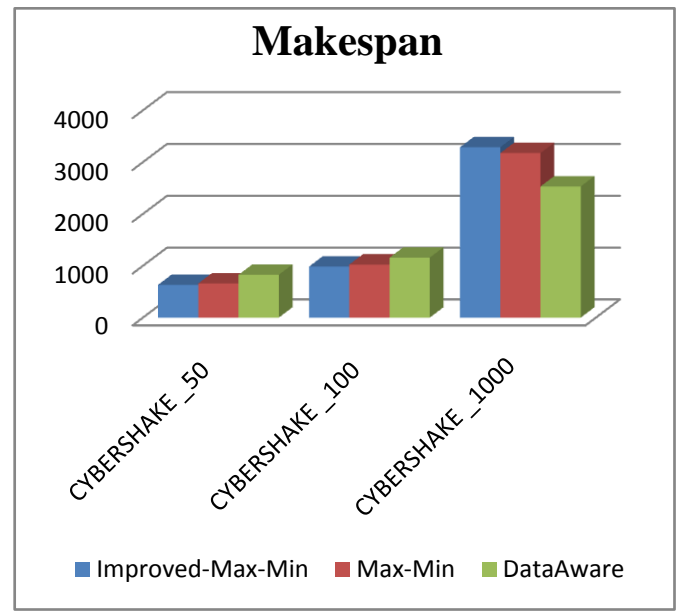

**Figure 6. Makespan comparison of different algorithms using 50, 100 and 1000 cloudlet of CyberShake**

Table 5 and Figure 6 shows and display the results of Inspiral dataset; It can be seen that, the improved Max-Min has been able to optimize completion time effectively than the other two algorithms by producing a better result when the number of cloudlets are 50 and 1000. It is only at cloudlet 100 that Max-Min is observed to produce slightly better results.

Similarly, the results obtained using Cybershake dataset were shown and display in Table 5 and Figure 6. It was observed that the improved Max-Min performed better when the number of cloudlet are 50 and 100, but fails to produce better results when the number of cloudlet is 1000. Thus, in general, our improved algorithm produces lower makespan than the traditional Max-Min and DataAware.

## 5. CONCLUSION AND FUTURE WORK

The paper presented an improved Max-Min algorithm with the focus of scheduling multiple cloudlet for improving makespan and efficient resource allocation. The algorithm is implemented using clooudsim. The results of the algorithm are compared with some other existing algorithms such as Max-Min and DataAware. The Improved algorithm has a better makespan than Max-Min and DataAware.

For future work, it is proposed that the research be extended to study the phenomenon of simultaneous executing tasks with maximum execution time and minimum execution time. This has the potential to avoid the issue of giving priority to tasks with maximum execution time over tasks with minimum execution time to inject efficiency when using Max-Min to schedule tasks in the cloud environment.

## 7. AUTHOR PROFILE

**James Kok Konjaang**
He is a senior instructor at Bolgatanga Polytechnic with more than five years teaching experience. His research interest includes Cloud Computing, Distributed Computing and Grid Computing. Received his HND in Marketing from Bolgatanga Polytechnic and BSc degree in Management with Computing from Regent University College of Science and Technology, both in Ghana in 2006 and 2012 respectively. He is currently pursuing a Master degree program in Distributed Computing from University Putra Malaysia.

**Jamilu Yahaya Maipan-uku**
B.Sc. Computer Science @ Ibrahim Badamasi Babangida University Lapai (IBBUL), Nigeria.
M.Sc. Computer Networks @ Universiti Putra Malaysia (UPM). Malaysia.
Grid Computing, Cloud Computing, Scheduling

**Kubuga Kennedy Kumangkem (Ph.D (IT Managment),**
MBA (IT Management), BSc (Computer Sci), Dip (Basic Edu)) Information Systems, ICT4D, Web Applications